\documentclass[10pt]{article}

\usepackage[pdftex]{graphicx}
\graphicspath{{./pdf/}}

\usepackage{csagh}
\usepackage{amsmath}
\usepackage{amsthm}
\usepackage{amssymb}

\usepackage{algorithm}
\usepackage{algorithmicx}
\usepackage{algpseudocode}

\usepackage{geometry}

\usepackage{times}

\begin{document}
\begin{opening}

\title{Improving PET scanner Time-Of-Flight Resolution using additional prompt photon}
\author[Department of Complex Systems, National Centre for Nuclear Research, 05-400 Otwock-Swierk, Poland, lech.raczynski@ncbj.gov.pl]{Lech Raczyński}
\author[High Energy Physics Division, National Centre for Nuclear Research, 05-400 Otwock-Swierk, Poland, wojciech.krzemien@ncbj.gov.pl]{Wojciech Krzemień}
\author[Department of Complex Systems, National Centre for Nuclear Research, 05-400 Otwock-Swierk, Poland, konrad.klimaszewski@ncbj.gov.pl]{Konrad Klimaszewski}

\begin{abstract}
Positronium Imaging requires two classes of events: double-coincidences originated from pair of back-to-back annihilation photons and triple-coincidences comprised with two annihilation photons and one additional prompt photon. 
The standard reconstruction of the emission position along the line-of-response of triple-coincidence event is the same as in the case of double-coincidence event; an information introduced by the high-energetic prompt photon is ignored. 
In this study, we propose to extend the reconstruction of position of triple-coincidence event by taking into account the time and position of prompt photon. 
We incorporate the knowledge about the positronium lifetime distribution and discuss the limitations of the method based on the simulation data. 
We highlight that the uncertainty of the estimate provided by prompt photon alone is much higher than the standard deviation estimated based on two annihilation photons. 
We finally demonstrate the extent of resolution improvement that can be obtained when estimated using three photons.
\end{abstract}


\end{opening}

\section{Introduction}

Positron emission tomography (PET) is a functional imaging method that is widely used in clinical oncology~\cite{Karp, Slomka, Conti}. 
PET is used for diagnosis, staging, and monitoring treatment of cancer. The leading tracer is 18-F-ﬂuorodeoxy-D-glucose (18-F-FDG) 
, which helps to label tissues with high glucose uptake, such as brain, liver, kidneys, and most cancers.
Together with other tomography techniques delivering morphologic and anatomic information, PET notablty improves the medical diagnosis power.

State-of-the-art PET detectors focus on estimating the spatial distribution of positron emitters by measuring the two annihilation photons.
These two anti-collinear photons are recorded by the pair of detectors mounted on one or more rings surrounding the patient. 
Detection of such event allows to identify the line-of-response (LOR) including the annihilation point. 
In addition to information about the LOR position, PET scanners provides also the time stamps of the arrival of the photons at the detectors. 
Thus for each event, time and position of annihilation may be estimated.
The acquisition of a large number of LORs, typically several millions, makes it possible to reconstruct the distribution of the radiotracer injected to the patient. 

However, PET may be also employed to the investigations of the local structure of the tissue via positron annihilation lifetime spectroscopy (PALS)~\cite{PALS1, PALS2}.
The positron history before the annihilation with an electron, not taken into consideration in the conventional PET measurement,  provides the information about the  tissue environment 
and can be useful in building the knowledge about cancer progression~\cite{Shibu2020,Science21,Steinberger24,Science24}.
It should be stressed that the positron-electron annihilation may proceed directly or via formation of positronium and the latter case occures in tissue with probability of about 40$\%$~\cite{Bass2023}.
Positronium can exist in two states, para-positronium (p-Ps) and ortho-positronium (o-Ps)~\cite{Harpen2003}. 
A short-lived p-Ps is formed in 25$\%$ of cases and a long-lived o-Ps in the remaining 75$\%.$ 
The main decay mode of p-Ps is into two photons while the o-Ps decays mostly into three photons. 
The self-annihilation mean lifetimes of p-Ps and o-Ps in vacuum are 125~ps and 142~ns, respectively~\cite{Cassidy}. 
However, in tissue, the mean lifetime of the o-Ps is reduced to about 2000~ps~\cite{Shibu2020,Jasi2017} 
and the decay results in two photons travelling in opposite directions along the LOR, exactly as in direct annihilation.
The mean value of o-Ps lifetime is sensitive to the local structure of the tissue.
The relationship between the mean o-Ps lifetime and the size of molecular voids (pores) is the basis of the PALS~\cite{Zgardz2020}; the shortening of the mean lifetime can be translated into the radius of pore using for example Tao-Eldrup model~\cite{Tao1972}.

Positronium lifetime  imaging is regarded as a novel biomarker that is independent to the distribution of the radiotracer~\cite{Science21,NIMA2024}. 
This measurement requires a special class of radionuclides that apart from the positron emits also a so-called high-energetic prompt photon;
lifetime of the positronium is estimated as the difference between the annihilation time and the time of the prompt photon emission.  
In this process it is assumed that prompt photon is emmited simultaneously in the moment of positronium formation.
Lifetime imaging in human tissues relies on the acquisition of triple-coincidences comprised of the pair of back-to-back annihilation photons and the prompt photon.

In previous positronium imaging studies, the  reconstruction of position of triple-coincidence event  is based only on the information coming from two annihilation photons traveling along the LOR~\cite{Shopa22,Shopa23,Shibu22,Jegal22,Qi22,Split24}.
In this paper, we propose to extend the position reconstruction algorithm of triple-coincidence event by taking into account additionally the time and position of prompt photon detection~\cite{PseudoTOF}. 
We will incorporate the knowledge about the positronium lifetime distribution to derive the algorithm for the position reconstruction. 
We will investigate the  improvement of  the position resolution using the additional prompt photon based on simulated PET data. 

The contributions of this work are twofold:  firstly, a method for reconstructing the position of triple-coincidence events is introduced and numerically validated.  To the author's best knowledge, this is the first time the high-energetic prompt photon was incorporated to the determination of position along the LOR in PET analysis. Secondly, the proposed method is compared with the reference reconstruction based on two annihilation photons only, highlighting the extent of resolution improvement obtained when estimated using three photons.

\section{Standard position reconstruction} \label{sec:stand_reco}

\begin{figure}[!ht]
\centering
\includegraphics[scale=.45]{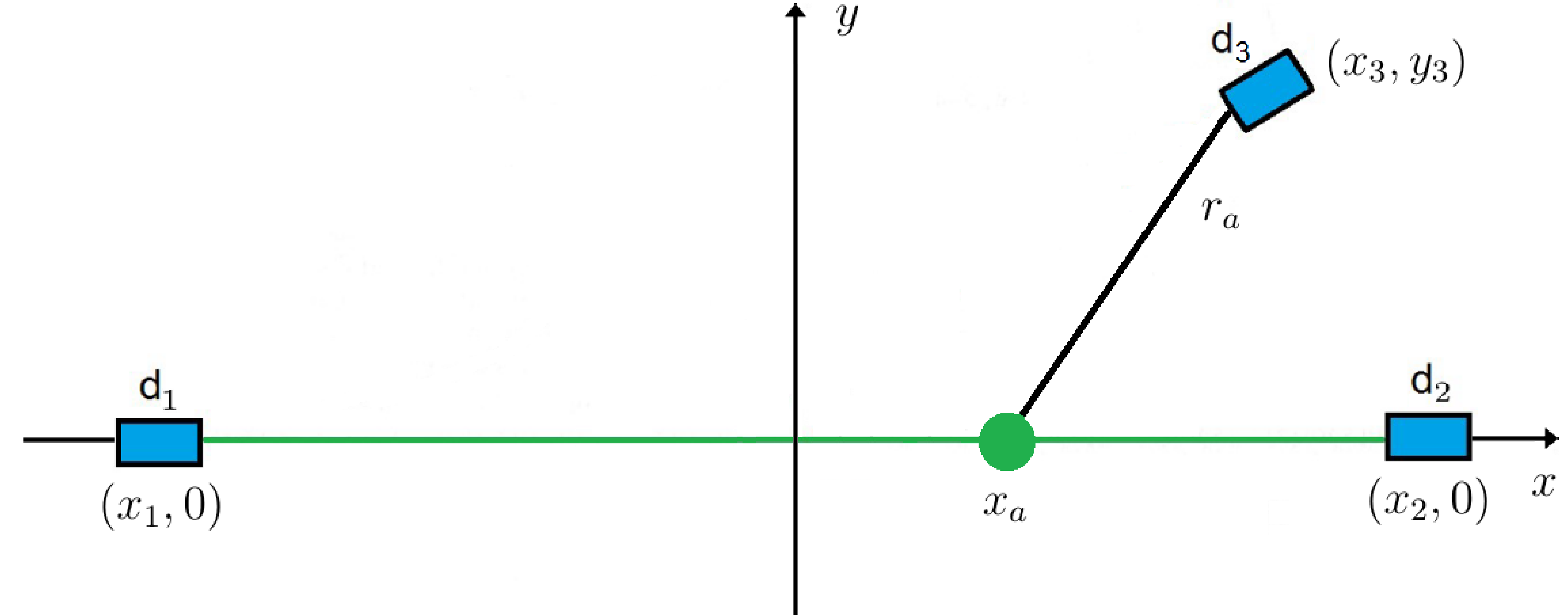}
\caption{Position reconstruction using two annihilation photons}
\label{rys:stand_reco}
\end{figure}

In this section we briefly describe the standard reconstruction of  the emission position of triple-coincidence.
State-of-the-art position reconstruction algorithm is based on times and positions of two annihilation photons only~\cite{Shopa22,Shopa23,Qi22,Split24}.
Without loss of generality, we may rotate and shift the original 3-dimensional coordinate system into 2-dimensional local space shown in Fig.~\ref{rys:stand_reco}; the LOR marked by two 
annihilation photons is a green horizontal line along the $x$ axis, and the detection positions of two annihilation photons are $(x_1, 0)$ and $(x_2, 0).$

In this algorithm the position of the prompt photon ($x_3, y_3$) is ignored and  the annihilation position along LOR ($x_a$), marked with green 
circle in Fig.~\ref{rys:stand_reco}, and time ($t_a$) is estimated using only the information given by two annihilation photons, i.e.,
\begin{align}
	t_a &= \frac{t_1 + t_2}{2} - \frac{|x_1-x_2|}{2c} \label{eq:ta} \\
	x_a &= \frac{c}{2} \left( t_1 - t_2 \right) \label{eq:xa}
\end{align}
where $t_1$ and $t_2$ are detection times in positions $x_1$ and $x_2,$ respectively, and $c$ is the  speed of light. 
In this work we consider only time uncertainties; times $t_1, t_2, t_3$ measured by detector $d_1, d_2, d_3$, respectively, are noisy and the positions of detections $(x_1,0), (x_2,0), (x_3,y_3) $, are known exactly.
We assume that time errors contributions are additive and normally distributed with standard deviation $\sigma_{t}.$
Therefore, the standard deviations of time $t_a$ and position $x_a$ calculated according to Eqs~\ref{eq:ta}-\ref{eq:xa} are:
\begin{align}
	\sigma_{t_a} &= \frac{\sqrt{2}}{2} \sigma_t \label{eq:sigma_ta} \\
	\sigma_{x_a} &= \frac{\sqrt{2}}{2} c\, \sigma_t. \label{eq:sigma_xa}
\end{align}

The goal of the standard position reconstruction of triple-coincidence is to provide the information sufficient to calculate the lifetime of the positronium, i.e., parameters  $x_a$ and $t_a$~\cite{Shopa22,Shopa23,Shibu22,Jegal22,Qi22,Split24}. 
Using Eqs~\ref{eq:ta}-\ref{eq:xa}, first distance $r_a$ travelled by the prompt photon from point $x_a$ to detector $d_3$ is evaluated (see Fig.~\ref{rys:stand_reco}):
\begin{equation}
	r_a = \sqrt{(x_a-x_3)^2 + y_3^2}  \label{eq:ra}
\end{equation}
and then the positronium lifetime $\tau$ is estimated as:
\begin{equation}
	\tau = t_a - t_p = t_a - t_3 + \frac{r_a}{c} \label{eq:tau}
\end{equation}
where $t_p$ and $t_3$ are times  of the prompt photon emission and detection, respectively.

We wish to make one comment about the uncertainties of detection times and positions. The positional uncertainty is limited to the size of the crystal and is typically of the order of a few mm. For time resolutions of state-of-the-art PET detectors, the standard deviation $\sigma_t$ dominates over the standard deviation of detection position normalized by the speed of light (see Eq.~\ref{eq:ta}). 
Therefore, in the first approximation, uncertainty along the positional dimension may be neglected during the calculations.

\section{Position reconstruction based on three photons}  \label{sec:extend_reco}

\begin{figure}[H]
\centering
\includegraphics[scale=0.45]{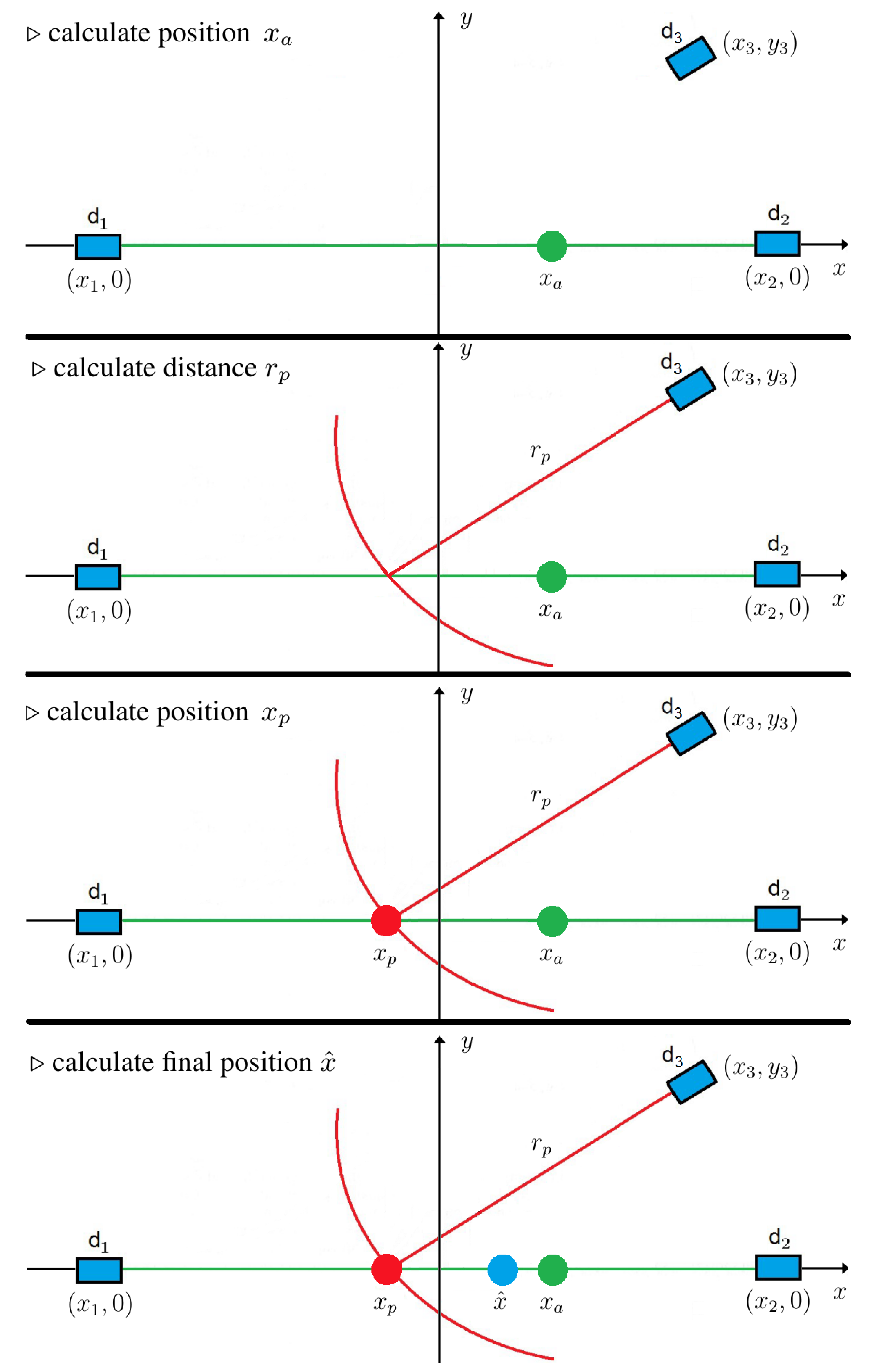}
\caption{Position reconstruction using three photons}
\label{rys:extend_reco}
\end{figure}

The idea of the extended position reconstruction algorithm is to improve the uncertainty $\sigma_{x_a}$ related with the estimate $x_a$ by providing second estimate $x_p$ marked with red circle in Fig.~\ref{rys:extend_reco} using the prior information about prompt photon recorded in detector $d_3.$ 
The final emission position of triple-coincidence $\hat{x}$, marked with blue circle in Fig.~\ref{rys:extend_reco}, is found as an optimal combination of the two estimates $x_a$ and $x_p.$
In this section we show details of derivation of the estimate $x_p$ and next  $\hat{x}.$

In the standard reconstruction presented in section~\ref{sec:stand_reco} the lifetime $\tau$ (see Eq.~\ref{eq:tau}) is calculated using the distance $r_a$ (see Eq.~\ref{eq:ra}).
In the extended algorithm the order of calculations is reversed: the distance $r_p \neq r_a$ travelled by the prompt photon (see Fig.~\ref{rys:extend_reco})  is evaluated based on the information about the 
lifetime $\tau.$ 
For this purpose  the knowledge about prior distribution of the lifetime is required.

In recent publication~\cite{Science21} it was shown  that the positronium lifetime $\tau$ is a random variable with probability density function (pdf):  
\begin{equation}
	\tau \sim \sum_{k=1}^3 I_k \cdot \exp \left(\frac{1}{\lambda_k} \right). \label{eq:tau_pdf}
\end{equation}
The lifetime distribution can be represented as a three-component exponential model arising from: para-positronium annihilation (p-Ps), ortho-positronium annihilation (o-Ps) and
process of direct annihilation of the positron and the electron without producing positronium.
The values of lifetimes ($1/\lambda_k$) and intensities ($I_k$) for $k = 1, 2, 3$ are gathered in Table~\ref{tab:lifetimes} and are based on the results presented in~\cite{Science21}.
\begin{table}[!ht]
\centering
\caption{Lifetime components}
\label{tab:lifetimes}
  \begin{tabular}{|r|r|r|}
                                 \hline
    & Intensity    & Mean lifetime \\
    & ($I_k$)    & (1/$\lambda_k$) \\\hline
   Direct & 0.65   & 388~ps  \\\hline
   p-Ps & 0.15  & 125~ps \\\hline
   o-Ps & 0.20 & 2000~ps \\\hline
  \end{tabular}
\end{table}

The distance  $r_p$ travelled by the prompt photon estimated based on the prior distribution of $\tau$ is:
\begin{equation}
	r_p = c \, (t_3 - t_a + \mu_{\tau}) \label{eq:rp}
\end{equation}	
where $\mu_{\tau}$ is the expected value (mean value) of the lifetime distribution defined in Eq.~\ref{eq:tau_pdf}.
Note that in case of the proposed method, the value of the lifetime $\tau$ (see Eq.~\ref{eq:tau}) is not used during the evaluation of the distance $r_p$ in Eq.~\ref{eq:rp}; we use only the expected value $\mu_{\tau}$.
According to Eq.~\ref{eq:rp}, uncertainties related with calcuation of distance $r_p$ 
may be modelled as a three-component exponentally modifed gaussian (EMG) distributions;
random variable $r_p$ is defined as sum of two independent normal distributions ($t_3$ and $t_a$) and  three-component exponential pdf ($\mu_{\tau}$), i.e.,
\begin{equation}
	r_p \sim c \, \sum_{k=1}^3  I_k \cdot \text{EMG} \left(\frac{1}{\lambda_k}, \frac{3}{2}\sigma_{t}^2 \right). \label{eq:rp_pdf}
\end{equation}
In the following we will approximate the distribution of $r_p$ with Gaussian pdf with standard deviation $\sigma_r$.

The position $x_p$ is calculated as the intersection of the circle with center at $(x_3, y_3)$ and radius $r_p$ marked with red in Fig.~\ref{rys:extend_reco} and LOR marked with green.
The value of $r_p$ greater than absolute value of $y_3$ guarantees that circle intersects the LOR in two points. 
The $x_p$ is selected as the one whose distance from position $x_a$ is smaller, i.e.,
\begin{equation}
x_p = v(r_p) =  x_3 + \text{sign}\left(x_a - x_3 \right) \sqrt{r_p^2 - y_3^2}. \label{eq:xp_def}
\end{equation}
The standard deviation $\sigma_{x_p}$ of the reconstructed value $x_p$ is evaluated 
using the linear term of the Taylor series of the function $v(r_p)$ in  Eq.~\ref{eq:xp_def}:
\begin{equation}
	\sigma_{x_p} = \sigma_{r} \frac{r_p}{\sqrt{r_p^2 - y_3^2}}.  \label{eq:sigma_xp}
\end{equation}

Finally, the estimate $\hat{x}$ is evaluated as:
\begin{equation} 
 	\hat{x} = x_a\frac{\sigma_{x_p}^2}{\sigma_{x_a}^2+\sigma_{x_p}^2} +  x_p\frac{\sigma_{x_a}^2}{\sigma_{x_a}^2+\sigma_{x_p}^2}. \label{eq:optim}
\end{equation}	
Eq.~\ref{eq:optim} shows the optimal way of combing estimates $x_a$ and $x_p$ with standard deviations $\sigma_{x_a}$ and $\sigma_{x_p},$ respectively~\cite{Kalman1,Kalman2,Jacobs93}.
$\hat{x}$ is the linear estimate whose standard deviation
\begin{equation} 
  	\sigma_{\hat{x}} = 	\frac{\sigma_{x_a}\sigma_{x_p}}{\sqrt{\sigma_{x_a}^2+\sigma_{x_p}^2}}	\label{eq:sigma_xhat}
\end{equation} 
is less than that of any other linear combination of $x_a$ and $x_p.$  
If $\sigma_{x_p}$ were equal to $\sigma_{x_a},$ Eq.~\ref{eq:optim} says that the optimal estimate of position is simply the average of the $x_a$ and $x_p.$
On the other hand, if $\sigma_{x_p}$ were larger than $\sigma_{x_a},$ then Eq.~\ref{eq:optim} dictates weighting $x_a$ more heavily than $x_p$.

It should be stressed that in case the value of radius $r_p$ is close to $y_3$  (radius of the circle in Fig.~\ref{rys:extend_reco} would be close to perpendicular to horizontal line along $x$ axis), 
the $\sigma_{x_p}$ according to Eq.~\ref{eq:sigma_xp} goes to infinity, and the $\hat{x}$  according to Eq.~\ref{eq:optim} goes to $x_a.$
Therefore, calculation of estimate $\hat{x}$ is reasonable only if distance:
\begin{equation}
	r_p > |y_3| + \kappa_{\text{min}} \,  \sigma_{r}	\label{eq:rp_ineq}
\end{equation}
where $\kappa_{\text{min}} > 0$ is an additional margin. 

The optimization of the value of the margin $\kappa_{\text{min}}$ will be provided during the simulation study and presented in details in section~\ref{sec:optim_kappa}. 
For each  triple-coincidence event the parameter $\kappa$ may be calculated:
\begin{equation}
	\kappa = \frac{r_p - |y_3|}{\sigma_{r}} 	\label{eq:kappa_def}
\end{equation}
and the extended position reconstruction will be provided only if 
\begin{equation}
	\kappa \geq \kappa_{\text{min}}. 	\label{eq:kappa_ineq}
\end{equation}
Otherwise standard reconstruction of the event position using algorithm described in Section~\ref{sec:stand_reco} is carried out.
Note that the inequality in Eq.~\ref{eq:kappa_ineq} is equivalent to the condition in Eq.~\ref{eq:rp_ineq}. 
Pseudo-code of the extended algorithm of position reconstruction is presented in Algorithm~1.

\begin{algorithm} \label{alg:extend}
\caption{Extended position reconstruction method using three photons}
\begin{algorithmic}[1] 
\Require $x_1, x_2, x_3, y_3, t_1, t_2, t_3, \sigma_{x_a}, \sigma_{r}, \mu_{\tau}, \kappa_{\text{min}}$
\State $t_a \Leftarrow \frac{t_1 + t_2}{2} - \frac{|x_1-x_2|}{2c} $ \Comment{calculate  time using two annihilation  photons: see Eq.~\ref{eq:ta}} 
\State $x_a \Leftarrow \frac{c}{2} \left( t_1 - t_2 \right) $  \Comment{calculate  position using two annihilation photons: see Eq.~\ref{eq:xa}} 
\State $r_p \Leftarrow c \, (t_3 - t_a + \mu_{\tau})$ \Comment{calculate distance $r_p$ using prior lifetime pdf: see Eq.~\ref{eq:rp}}
\State $\kappa \Leftarrow \frac{r_p - |y_3|}{\sigma_{r}}$ \Comment{calculate parameter $\kappa$: see Eq.~\ref{eq:kappa_def}}
\If{$\kappa \geq \kappa_{\text{min}}$}
      \State $x_p \Leftarrow x_3 - \text{sign}\left(x_a - x_3 \right) \sqrt{r_p^2 - y_3^2}$ \Comment{calculate  position using prompt  photon alone: see Eq.~\ref{eq:xp_def}}
	\State $\sigma_{x_p} \Leftarrow \sigma_{r} \frac{r_p}{\sqrt{r_p^2 - y_3^2}}$ \Comment{calculate standard deviation of $x_p:$ see Eq.~\ref{eq:sigma_xp}}
      \State $\hat{x} \Leftarrow x_a\frac{\sigma_{x_p}^2}{\sigma_{x_a}^2+\sigma_{x_p}^2} +  x_p\frac{\sigma_{x_a}^2}{\sigma_{x_a}^2+\sigma_{x_p}^2}$ \Comment{calculate final position using three photons: see Eq.~\ref{eq:optim}}
	\State $\sigma_{\hat{x}} \Leftarrow 	\frac{\sigma_{x_a}\sigma_{x_p}}{\sqrt{\sigma_{x_a}^2+\sigma_{x_p}^2}}$ \Comment{calculate standard deviation of $\hat{x}:$ see Eq.~\ref{eq:sigma_xhat}}
\Else
	 \State $\hat{x} \Leftarrow x_a $	 \Comment{calculate final position using standard algorithm} 
	\State $\sigma_{\hat{x}} \Leftarrow 	\sigma_{x_a}$ \Comment{calculate standard deviation of $\hat{x}$ using standard algorithm}
\EndIf
\end{algorithmic}
\end{algorithm}

\section{Results}

\begin{figure}[!ht]
\centering
\includegraphics[scale=.8]{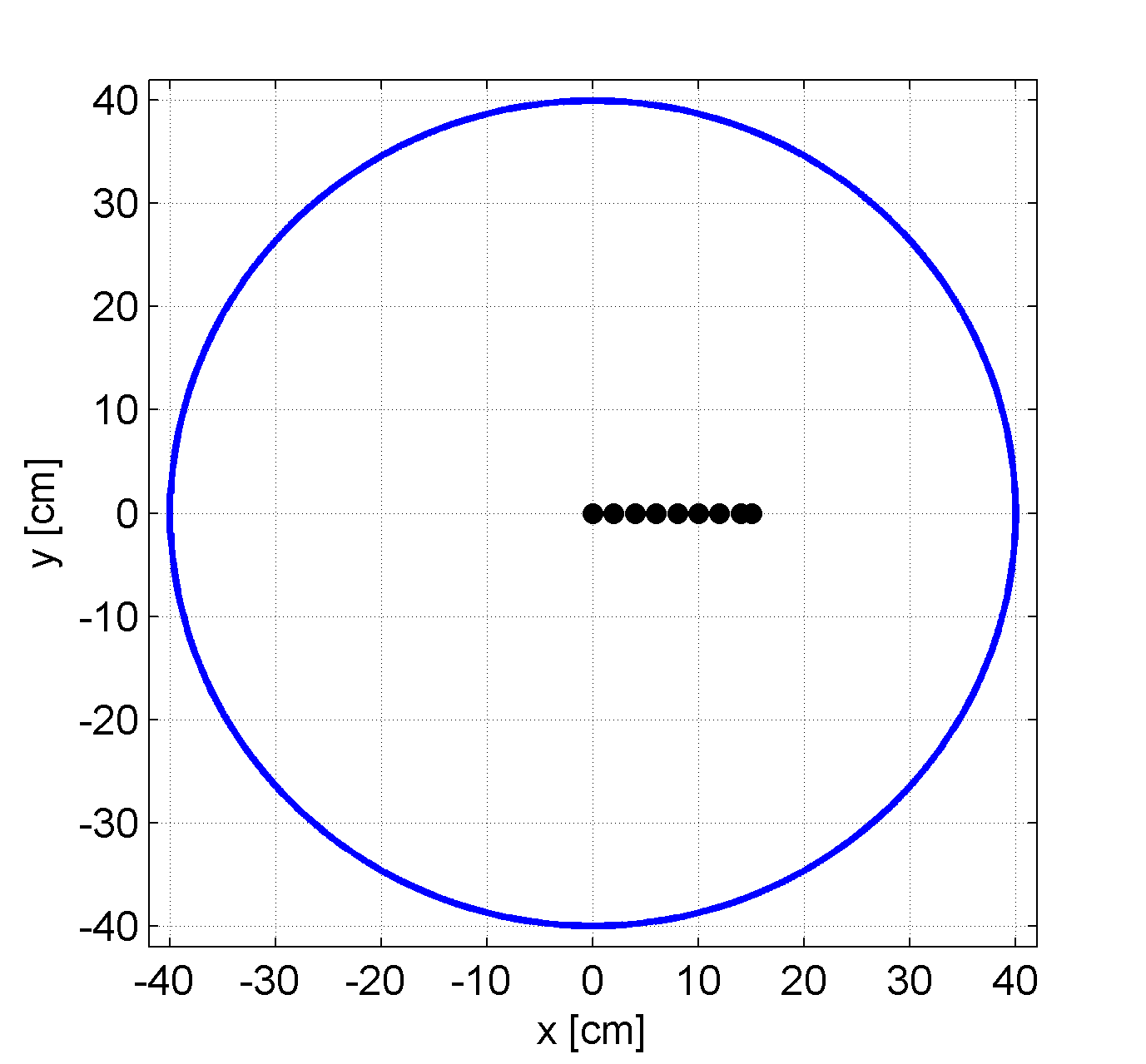}
\caption{Simulation setup for the measurement of triple-coincidence events}
\label{rys:scanner}
\end{figure}

In this section we demonstrate proof of concept of the  reconstruction algorithm proposed in section~\ref{sec:extend_reco}. 
Sample data were generated using Monte Carlo simulation prepared in MATLAB 7.14.0 (R2012a). 
We modeled 2-dimensional PET scanner geometry deﬁned as a infinitely thin cylinder with radius of 40~cm (see Fig.~\ref{rys:scanner}).
The simulated coincidence resolving time resolution of the scanner, denoted hereafter with $\text{CRT}_{\text{ref}},$  was 500~ps~\cite{Slomka,Sluis}.
This value is defined using full width at half maximum (FWHM) as:
\begin{equation}
	\text{CRT}_{\text{ref}} = \text{FWHM}\left(t_1 - t_2 \right) \label{eq:CRT_def}
\end{equation}
and corresponds to the standard deviation  of the single time measurement in detector 
\begin{equation}
	\sigma_t \approx \frac{\text{CRT}_{\text{ref}}}{\sqrt{2}\cdot 2.35} =  150~\text{ps}.\label{eq:sigma_t}
\end{equation}
We simulated point sources placed in nine different radial positions in the cylinder: 0, 2, 4, 6, 8, 10, 12, 15~cm  (see Fig.~\ref{rys:scanner}).
At each position we generated in total 100,000 triple-coincidence events and reconstructed the emission point using both standard and extended algorithms presented in previous sections.
In each case the value of time resolution was  estimated based on the standard deviation of the position reconstruction combining Eqs~\ref{eq:sigma_xa} and~\ref{eq:sigma_t}:
\begin{equation}
	\hat{\text{CRT}} = 2 \cdot 2.35\cdot \frac{\sigma_{\hat{x}}}{c}. \label{eq:CRT_new}
\end{equation}
From Eq.~\ref{eq:CRT_new} we receive immediately that $\sigma_{x_a}$ for standard position reconstruction ($\text{CRT}_{\text{ref}}$~= 500~ps) is:
\begin{equation}
	\sigma_{x_a}  = \frac{\text{CRT}_{\text{ref}} \cdot c}{2 \cdot 2.35} = 3.18~\text{cm.} \label{eq:sigma_xa_318}
\end{equation}

\subsection{Lower bound of CRT estimated using three photons} \label{sec:lower_bound}

In the first step we will derive  the value of the  lower bound of the time resolution of the PET detector  obtained with additional information provided by the prompt photon.
Substituting Eq.~\ref{eq:sigma_xhat} to Eq.~\ref{eq:CRT_new}, and knowing the relation between $\sigma_{x_a}$ and $\text{CRT}_{\text{ref}}$ given in Eq.~\ref{eq:sigma_xa_318}, we may write that:
\begin{equation}
	\hat{\text{CRT}} = \text{CRT}_{\text{ref}} \frac{\sigma_{x_p}}{\sqrt{\sigma_{x_a}^2+\sigma_{x_p}^2}}. \label{eq:CRT_CRTref}
\end{equation}
The reference value of $\text{CRT}_{\text{ref}}$ implies directly the value of standard deviation $\sigma_{x_a}$ (see Eq.~\ref{eq:sigma_xa_318}).
In the following we will show that the $\text{CRT}_{\text{ref}}$ also has an impact on $\sigma_{x_p}$ and we estimate its smallest value.
Using the linear approximation of $\sigma_{x_p}$ in Eq.~\ref{eq:sigma_xp} it may be shown that:
\begin{equation}
	\sigma_{x_p} = \sigma_{r} \frac{r_p}{\sqrt{r_p^2 - y_3^2}} \geq \sigma_{r}  \label{eq:ineq_sigma_xp}
\end{equation}
since 
\begin{equation}
	\frac{r_p}{\sqrt{r_p^2 - y_3^2}} \geq 1 \quad\quad \text{for any} \quad\quad r_p > y_3. \label{eq:ineq_1}
\end{equation}
Substituting inequality in Eq.~\ref{eq:ineq_sigma_xp} to Eq.~\ref{eq:CRT_CRTref} we get the lower bound of the CRT:
\begin{equation}
	\hat{\text{CRT}} \geq \text{CRT}_{\text{ref}} \frac{\sigma_{r}}{\sqrt{\sigma_{x_a}^2+\sigma_{r}^2}}. \label{eq:CRT_lower}
\end{equation}
The only parameter that is missing in Eq.~\ref{eq:CRT_lower} is the standard deviation $\sigma_{r}.$
The value of $\sigma_{r}$ that approximates the distribution of error of distance $r_p$ (see Eq.~\ref{eq:rp_pdf}) depends on $\text{CRT}_{\text{ref}}$ due to presence of the $\sigma_t$ parameter that describes uncertainties of registration times. 
In this study, the evaluation of the general function of $\sigma_{r}$ on $\sigma_t$ is not crucial, and we will estimate only the value of  $\sigma_{r}$ for  $\sigma_t$ specified for the considered PET detector 
(see Eq.~\ref{eq:sigma_t}).  

\begin{figure}[!ht]
\centering
\includegraphics[scale=.7]{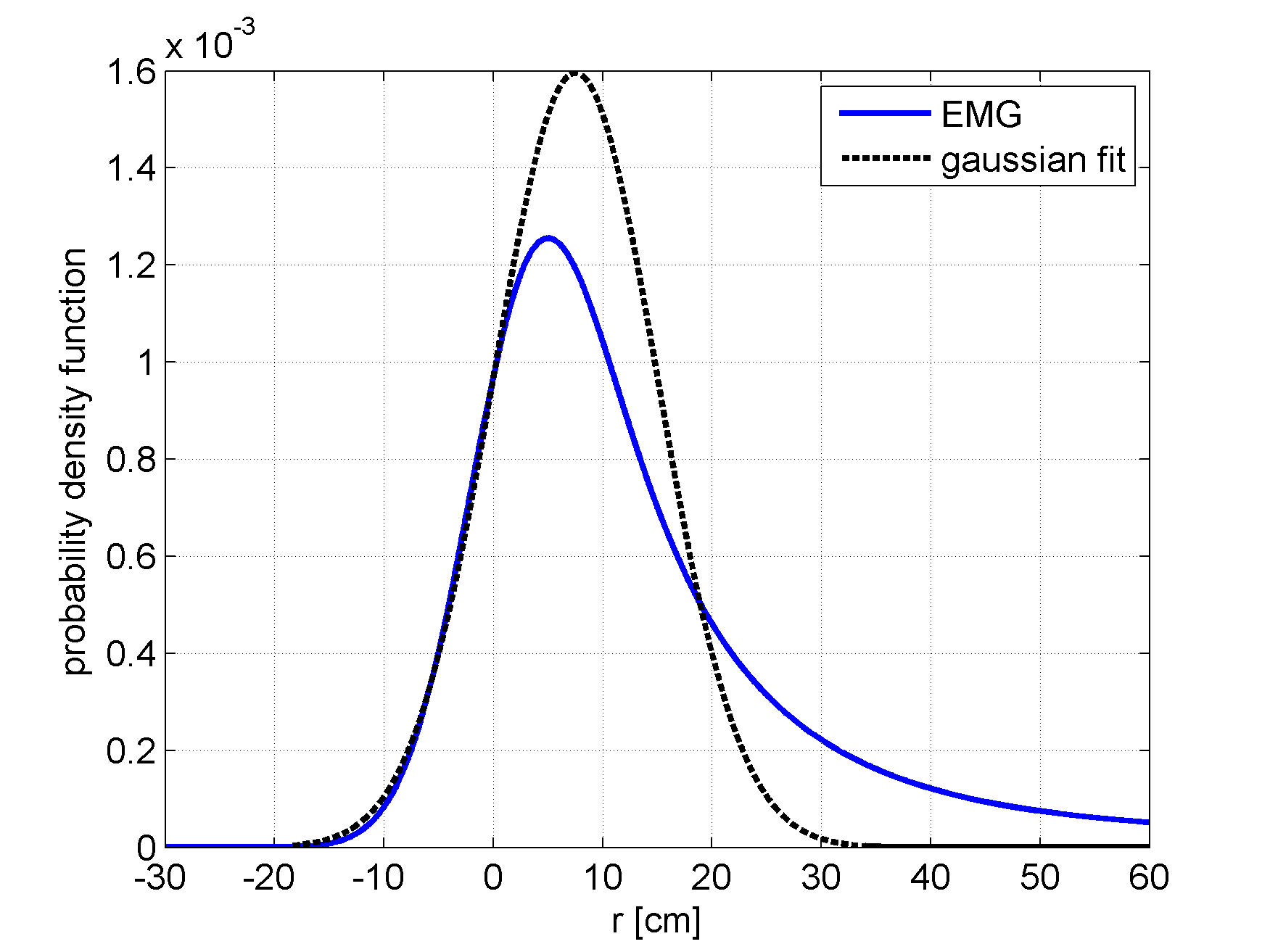}
\caption{Approximation of the three-component EMG distribution with Gaussian fit}
\label{rys:EMG_fit}
\end{figure}
Theoretical pdf function of distance $r_p$ modelled as a sum of three EMG functions according to Tab.~\ref{tab:lifetimes}, calculated for $\sigma_t$~=~150~ps, is marked with blue in Fig.~\ref{rys:EMG_fit}.
The EMG function was calculated under the assumption that expected values of variables $t_a$ and $t_3,$ given with normal distributions, are 0. 
It should be stressed that the EMG function is very sensitive to selected parameters of mean lifetimes and intensities (see Tab.~\ref{tab:lifetimes}). 
Therefore, a more general assumption about the shape of the distribution is required and in this study, a Gaussian fit was chosen. 
Approximation of the EMG function with normal distribution  is marked with black dotted curve in Fig.~\ref{rys:EMG_fit}.
The parameters of the Gaussian fit are  $\mu_{r}$~=~7.5~cm and $\sigma_{r}$~=~7.5~cm.
The mean value of the Gaussian function $\mu_r$ corresponds to the mean value $\mu_{\tau}$ of the lifetime distribution  $\tau$ defined  in Eq.~\ref{eq:tau_pdf}.
According to Eqs~\ref{eq:rp} and~\ref{eq:rp_pdf} we may write that:
\begin{equation}
	\mu_{\tau} = \frac{\mu_{r}}{c} = 250~\text{ps}. \label{eq:mu_tau_250ps}
\end{equation}

Substituting the value  $\sigma_{r}$~=~7.5~cm and previous value of $\sigma_{x_a}$~=~3.18~cm (see Eq.~\ref{eq:sigma_xa_318}) to the inequality in Eq.~\ref{eq:CRT_lower} we get finally the lower bound of the CRT:
\begin{equation}
	\hat{\text{CRT}} \geq 460~\text{ps}. \label{eq:CRT_lower_final}
\end{equation}
The lower bound of the time resolution estimated using three photons is about 8$\%$ smaller that the reference value 500~ps of CRT evaluated using only two annihilation photons.

It should be stressed that there are two reasons why the lower bound CRT of about 460~ps cannot be achieved in real experiments for systems with reference 500~ps time resolution.
First of all, the extended algorithm proposed in section~\ref{sec:extend_reco} may be applied only if the condition using parameter $\kappa$ in Eq.~\ref{eq:kappa_ineq} is fulfilled (see line 5 in pseudo-code in Algorithm 1).
Secondly, the experimental distribution of the standard deviation $\sigma_{x_p}$ should be taken into account (see Eq.~\ref{eq:CRT_CRTref}); in this section, only the smallest value of  
$\sigma_{x_p},$ i.e., $\sigma_{r}$ (see Eq.~\ref{eq:CRT_lower}), was analyzed. 
In the next step, we will derive the effective value of the CRT by considering the two remarks mentioned above and we will show that the main parameter that has the impact on the 
performance of the proposed algorithm is $\kappa_{\text{min}}$.

\subsection{Optimization of the reconstruction parameter $\kappa_{\text{min}}$} \label{sec:optim_kappa}

In this section, we will describe the procedure of optimization of the  $\kappa_{\text{min}}$ parameter and we will derive the effective value of the CRT. 
For this purpose, we will use   the simulation data for the point source in the center of the PET scanner.
First, in subsection~\ref{sec:optim_kappa_1} we will derive the experimental distribution of the $\kappa$ parameter.
Next, in subsection~\ref{sec:optim_kappa_2} we will introduce the dependence of the CRT on the $\kappa$ parameter.
Finally, in subsection~\ref{sec:optim_kappa_3} we will show that the $\kappa$ parameter trades off between the number of the events considered in the proposed algorithm and the accuracy of the reconstruction. 

\subsubsection{Derivation of the cumulative distribution function of the $\kappa$ parameter} \label{sec:optim_kappa_1}

\begin{figure}[!ht]
\centering
\includegraphics[scale=.7]{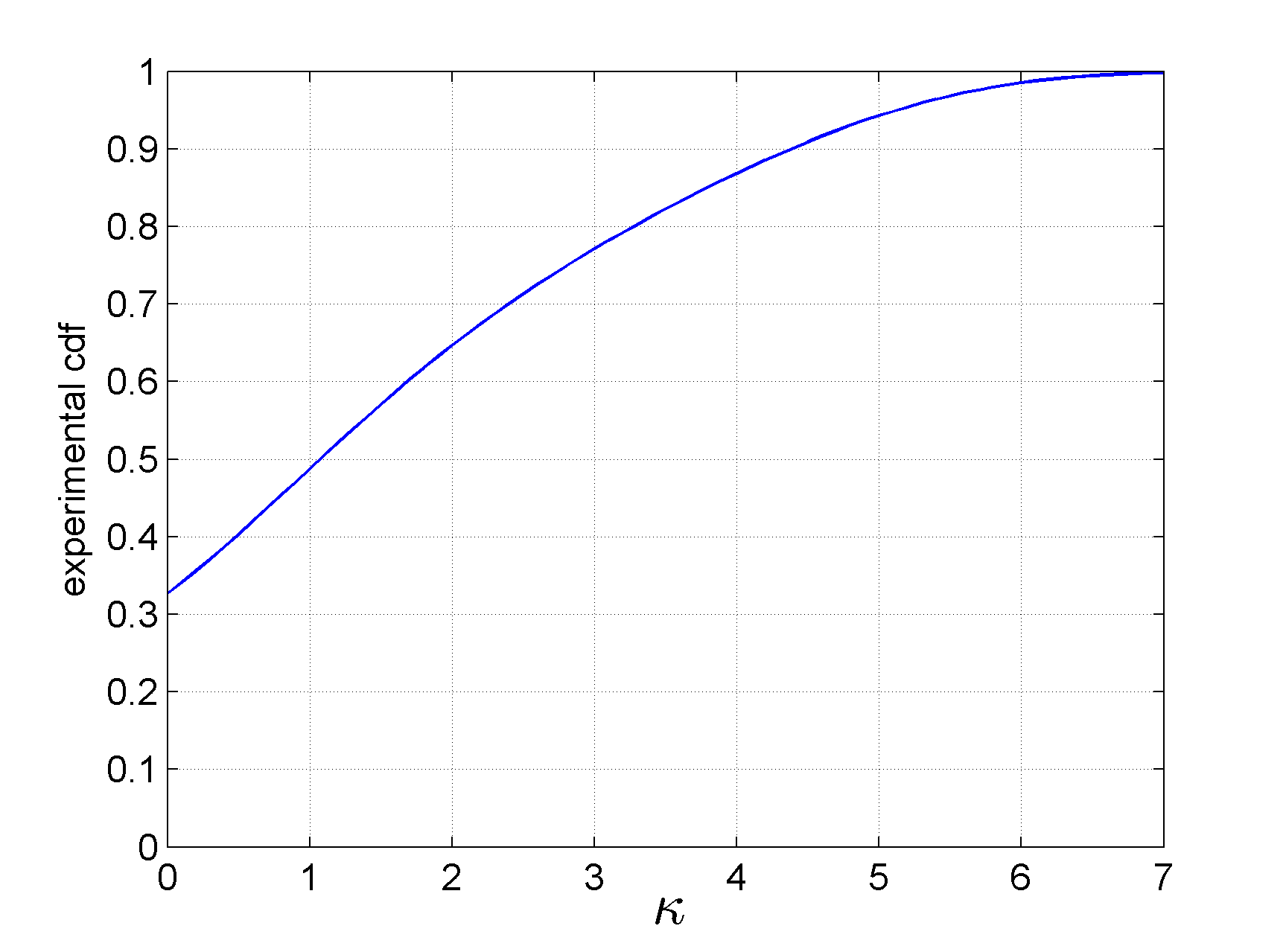}
\caption{Cumulative distribution function of the $\kappa$ parameter of the simulation data for the point source in the center of the PET detector}
\label{rys:kappa}
\end{figure}

Using the simulation data with 100,000 triple-coincidence events we first evaluated the experimental  cumulative distribution function (cdf) of the $\kappa$ parameter (see Eq.~\ref{eq:kappa_def}). 
The resulting cdf function is shown in Fig.~\ref{rys:kappa}.
One may see that for about 30$\%$ of events the reconstruction using the extended method is impossible since  $\kappa \leq 0.$
From Eq.~\ref{eq:rp_ineq} in that case the distance $r_p$ is smaller than detection position $y_3;$ circle with radius $r_p$ has no intersections with a line marked by two annihilation photons (see Fig.~\ref{rys:extend_reco}). Hence, in at least  30$\%$ of cases the standard position reconstruction has to be applied.
The experimental cdf function for $\kappa$ smaller than 0 was not presented in Fig.~\ref{rys:kappa} as these values of parameter have no influence on the resulting performance of the reconstruction algorithm;
in the further analysis we will investigate only positive values of the $\kappa$ parameter.
The highest value of $\kappa$ observed in the simulation data was about 7; for this value cdf converges to 1 (see  Fig.~\ref{rys:kappa}). 

\subsubsection{Investigation of dependence of the CRT on the $\kappa$ parameter} \label{sec:optim_kappa_2}

\begin{figure}[!ht]
\centering
\includegraphics[scale=.7]{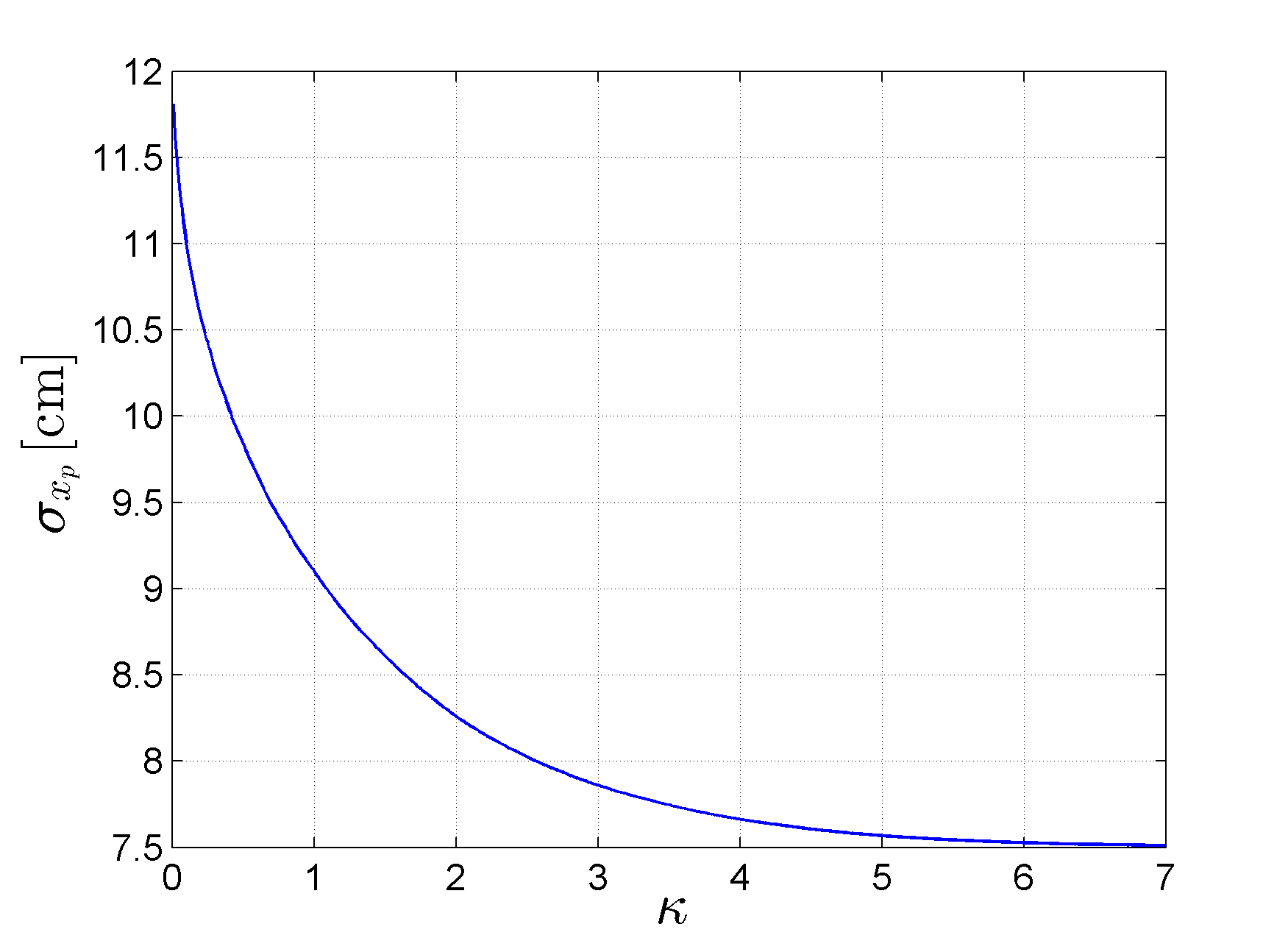}
\caption{Mean value of the standard deviation $\sigma_{x_p}$ after selecting the data based on the parameter $\kappa$}
\label{rys:kappa_vs_sigmap}
\end{figure}

During the optimization of the $\kappa$ parameter one needs to take into account that for the specified $\kappa$ the mean value of the standard deviation  $\sigma_{x_p}$ changes. 
The mean value of $\sigma_{x_p}$ as a function of the  parameter $\kappa$ is presented in Fig.~\ref{rys:kappa_vs_sigmap}.
For a given event with the smallest considered $\kappa$ close to 0, according to Eq.~\ref{eq:sigma_xp}, the standard deviation $\sigma_{x_p}$ goes to infinity.
Therefore, in the dataset with the events with $\kappa  \geq$~0 the mean value of the  $\sigma_{x_p}$ is the largest and is about 12~cm (see Fig.~\ref{rys:kappa_vs_sigmap}).
Note that in that case, the largest amount of triple-coincidence events, of about 70$\%,$ may be used to position reconstruction based on extended algorithm 
(see the cdf function in corresponding region for $\kappa \approx$~0 in Fig.~\ref{rys:kappa}).
Increasing the minimal value of $\kappa$ from 0 to 7, one improves the mean value of  standard deviation $\sigma_{x_p}$ by removing the events with the worst reconstruction; the function in Fig.~\ref{rys:kappa_vs_sigmap} is decreasing. 
It should be stressed that for $\kappa \approx 7$  the mean standard deviation $\sigma_{x_p}$ converges to the value $\sigma_{r}$~=~7.5~cm; the smallest possible value of standard deviation according to  Eq.~\ref{eq:ineq_sigma_xp}.
However, at the same time the number of the selected events for which the proposed method may be applied also reduces  and 
for $\kappa \approx$~7 this number goes to 0 (see the cdf function in Fig.~\ref{rys:kappa}).
For this reason we observe the trade-off between the mean value of  standard deviation $\sigma_{x_p}$ and the number of triple-coincidences for which the condition using parameter $\kappa$ in Eq.~\ref{eq:rp_ineq} is met. 

\begin{figure}[!ht]
\centering
\includegraphics[scale=.7]{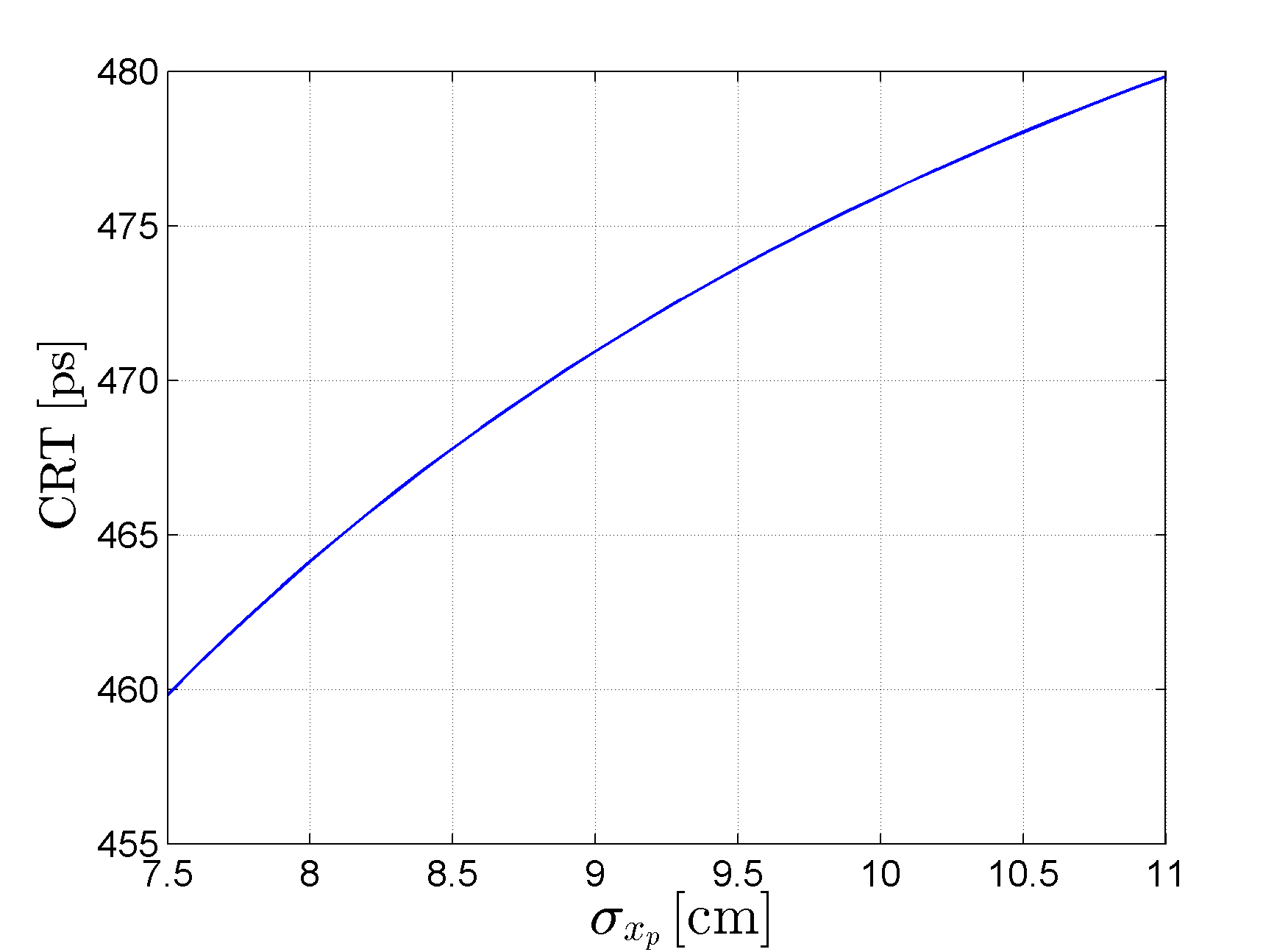}
\caption{Dependence of the $\hat{\text{CRT}}$ on the standard deviation $\sigma_{x_p}$}
\label{rys:CRT_hat}
\end{figure}

In Fig.~\ref{rys:CRT_hat} the theoretical dependence of the scanner CRT on the mean value of the  standard deviation $\sigma_{x_p}$ is shown (see Eq.~\ref{eq:CRT_CRTref}).
Note that the smallest value $\hat{\text{CRT}}$ of about 460~ps corresponds to the lower bound estimated in previous section (see Eq.~\ref{eq:CRT_lower_final}) and is estimated for 
$\sigma_{x_p} = \sigma_{r} =$~7.5~cm.

Taking into account that a given value of $\sigma_{x_p}$ corresponds to the specified threshold of $\kappa,$ as it was presented in Fig.~\ref{rys:kappa_vs_sigmap},
we have to expect that the effective value of CRT, denoted hereafter with  $\text{CRT}_{\text{eff}},$ will be larger than the value $\hat{\text{CRT}}$ illustrated in  Fig.~\ref{rys:CRT_hat}.
For instance, if the threshold of the parameter $\kappa$ was set to 3, from Fig.~\ref{rys:kappa_vs_sigmap} we may expect that  the mean value of  the standard deviation $\sigma_{x_p}$ in the dataset is about 7.8~cm. 
For this value of $\sigma_{x_p},$  from Fig.~\ref{rys:CRT_hat} we see immediately that the $\hat{\text{CRT}}$ is about 463~ps.
However, this value is not the final CRT of the PET system since from  Fig.~\ref{rys:kappa} we may see that selection of only the events with $\kappa \geq $~3 implies that for about 77$\%$ of events  the standard position reconstruction had to be applied; the effective CRT is larger than the theoretical value presented in  Fig.~\ref{rys:CRT_hat}. For a distinct  value of the $\kappa$ parameter used for selection criteria, a different effective value of the time resolution will be obtained.

\subsubsection{Final optimization of the reconstruction parameter $\kappa_{\text{min}}$} \label{sec:optim_kappa_3}

The problem of  finding the smallest effective resolution  time  ($\text{CRT}_{\text{eff}}$) is equivalent to the optimization task of the  $\kappa$ parameter.
For this purpose, we need to recall the definition of the standard deviation calculated using two reconstructions $x_a$ (standard algorithm see Eq.~\ref{eq:xa}) and $\hat{x}$ 
(extended algorithm see Eq.~\ref{eq:optim}):
\begin{equation}
	\sigma_{x_{\text{eff}}}^2 = \frac{1}{N_{a}+\hat{N}} \left( \sum_i^{N_a} \left(x_a(i) - m \right)^2 + \sum_j^{\hat{N}} \left(\hat{x}(j) - m \right)^2 \right). \label{eq:sigma_xeff}
\end{equation}
In Eq.~\ref{eq:sigma_xeff} values $N_a$ and $\hat{N}$ denote the numbers of the triple-coincidence events for which the  standard  and the extended reconstruction methods were applied, respectively.
In other words, in $N_a$ cases the calculated value of $\kappa$ was smaller than the selected threshold and for the remaining $\hat{N}$ cases the parameter $\kappa$ was higher than the threshold.
We assumed that both methods are unbiased, and the same mean value of the position reconstruction $m$ is obtained.
Taking into account that:
\begin{align} 
	\sigma_{x_a}^2 &= \frac{1}{N_a} \sum_i^{N_a} \left(x_a(i) - m \right)^2 \label{eq:siga_def} \\
	\sigma_{\hat{x}}^2 &= \frac{1}{\hat{N}} \sum_j^{\hat{N}} \left(\hat{x}(j) - m \right)^2 \label{eq:sighat_def}
\end{align}	
based on Eq.~\ref{eq:sigma_xeff} we may write that:
\begin{equation}
	\sigma_{x_{\text{eff}}}^2 = \frac{1}{N_{a}+\hat{N}} \left( N_a \,\sigma_{x_a}^2 + \hat{N} \,\sigma_{\hat{x}}^2 \right). \label{eq:sigma_xeff_v2}
\end{equation}
The effective value of the CRT calculated using Eq.~\ref{eq:sigma_xeff_v2} and based on previous derivations of $\sigma_{x_a}$ and $\sigma_{\hat{x}}$ in Eqs~\ref{eq:sigma_xa_318} and~\ref{eq:CRT_new}, respectively, is given as:
\begin{equation}
	\text{CRT}_{\text{eff}} = \sqrt{\frac{N_{a}}{N_{a}+\hat{N}} \,\text{CRT}_{\text{ref}}^2 + \frac{\hat{N}}{N_{a}+\hat{N}} \,\hat{\text{CRT}^2}}. \label{eq:CRT_xeff}
\end{equation}

\begin{figure}[!ht]
\centering
\includegraphics[scale=.7]{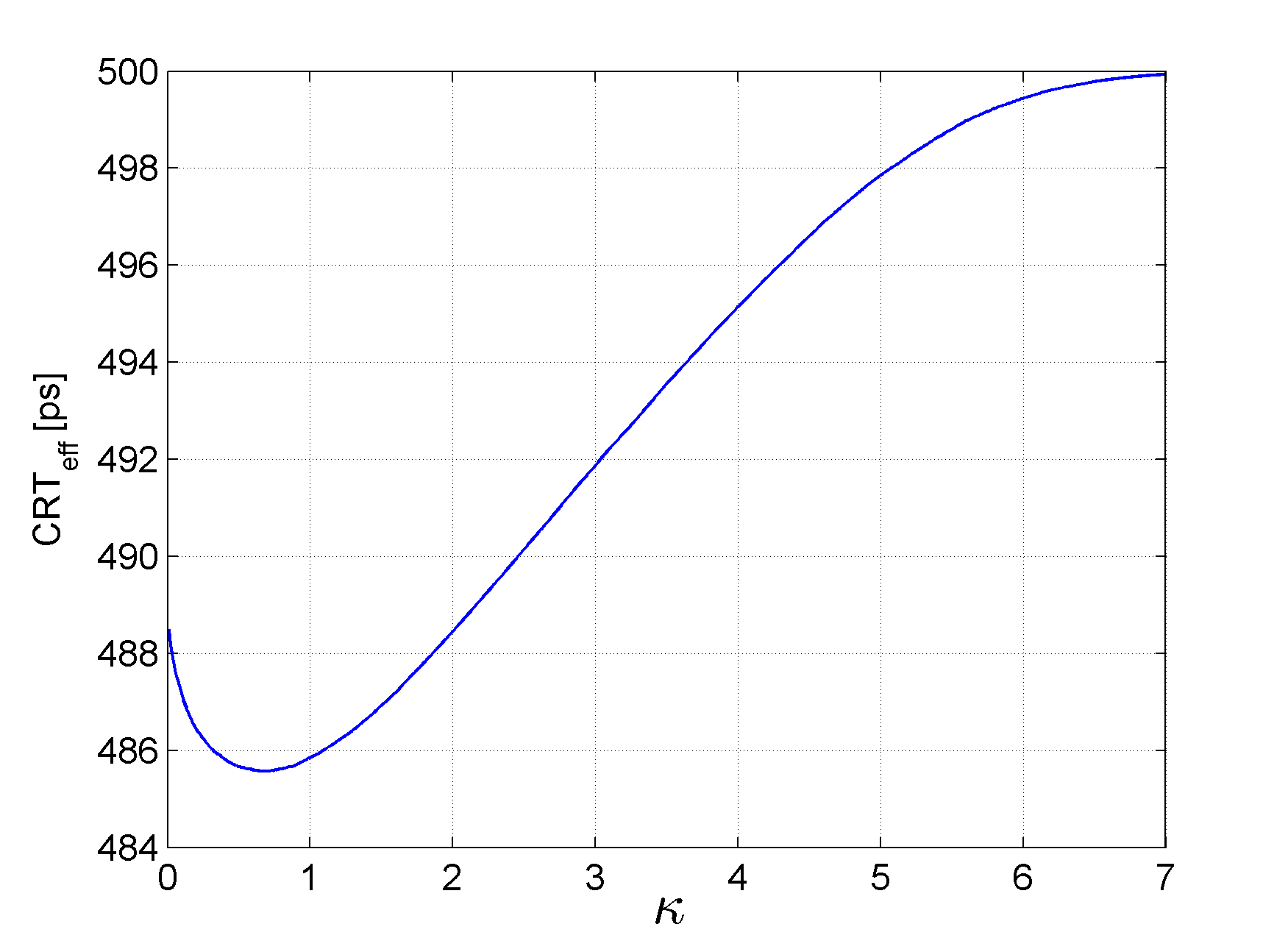}
\caption{Calculation of the effective CRT as the function of $\kappa$ parameter}
\label{rys:CRT_effect_theory}
\end{figure}

Fig.~\ref{rys:CRT_effect_theory} presents the relation between the $\text{CRT}_{\text{eff}}$ and the value of the selected $\kappa$ parameter.
The parameter $\kappa$ trades off the number of the events considered in the proposed algorithm and the accuracy of the reconstruction.
Large values of the parameter favor triple-coincidence events with best position resolution. For $\kappa$ about 7 no events meet the condition in Eq.~\ref{eq:rp_ineq} and the effective CRT converges to the reference value of time resolution of 500~ps. 
Decreasing the value of $\kappa$ tends to improve the performace of the reconstruction.
The smallest  $\text{CRT}_{\text{eff}} \approx$~486~ps  is observed for $\kappa$ about 0.7. For smaller values of $\kappa$ the time resolution starts to increase.

\subsection{Simulation study of estimation of CRT using three photons}

In the last part of this study, we will derive the effective value of the CRT in different radial positions of the PET scanner based on the simulation data. 
We will also compare the values of CRT estimated using the proposed scheme and the reference method. 
During the comparative study presented  in this section  we applied the threshold $\kappa_{\text{min}} =$~0.7 based on the results shown in Fig.~\ref{rys:CRT_effect_theory}.
This  requirement imposes that the extended position reconstruction method uses about 55$\%$ of acquired triple-coincidence events (see Fig.~\ref{rys:kappa}).
For the estimation of the position in remaining 45$\%$ of cases  only the information from two annihilation photons was taken into account; 
in that case the resulting positions for both compared methods  are the same. 

\begin{figure}[!ht]
\centering
\includegraphics[scale=.7]{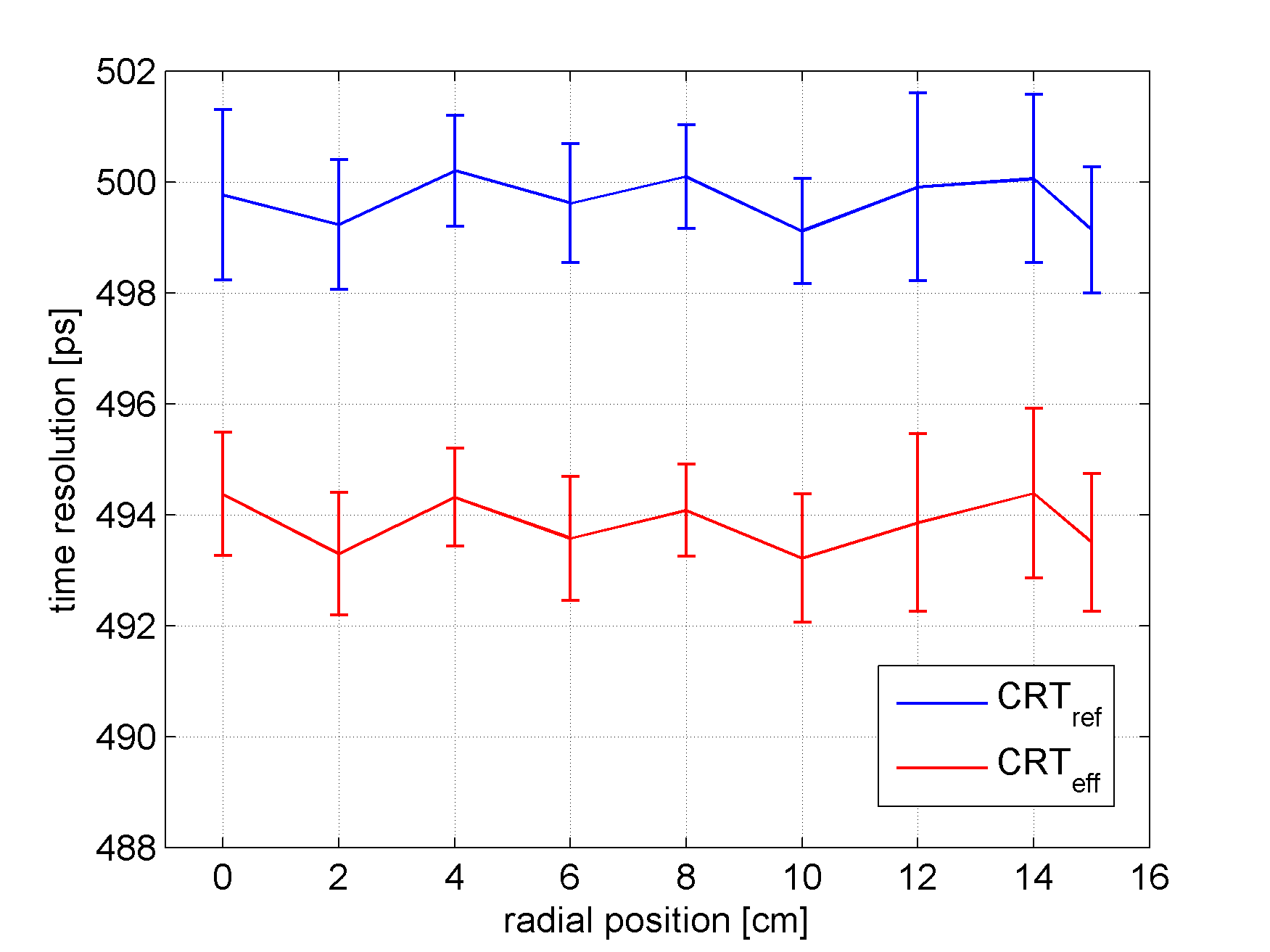}
\caption{Calculation of the CRT using two reconstruction methods: stanrdard (blue curve), extended (red curve)}
\label{rys:CRT_final}
\end{figure}

Fig.~\ref{rys:CRT_final}  compares the time resolutions of the PET scanner calculated in different radial positions using two methods: reference reconstruction algorithm (blue curve) and proposed 
reconstruction algorithm (red curve).
The error bars indicate standard deviations and were estimated from the 10 realizations of event smearing with assumed resolution of 500~ps.
The evaluation of the distribution of the CRT for the reference method ($\text{CRT}_{\text{ref}}$) was performed only for validation purposes; 
as expected the resulting curve is stable at the level of 500~ps. 
The effective value of the time resolution estimated using the proposed scheme  ($\text{CRT}_{\text{eff}}$) is on average about 494~ps and is slightly better than the $\text{CRT}_{\text{ref}}$. 
The  experimental value of $\text{CRT}_{\text{eff}}$ for center position is about 8~ps higher than the theoretical one for $\kappa_{\text{min}} =$~0.7 shown in Fig.~\ref{rys:CRT_effect_theory}.

\section{Conclusions}

In this article, the extended  method  for position reconstruction of the triple-coincidence event was  introduced.
The key feature of this model is the incorporation of knowledge about  the time and position of prompt photon detection. 
We have evaluated the lower bound of the coincidence resolving time resolution for the proposed method and we have shown that this value is about 8$\%$ smaller in comparison to the reference scanner CRT of 500~ps.
We highlighted the statistical phenomenon that deteriorates the quality of the reconstruction: 
the uncertainty of the estimate provided by prompt photon alone ($\sigma_{x_p}$) is much higher than the standard deviation of the reference model ($\sigma_{x_a}$).
This leads to (1) decreasing the quality of the estimate based on both reconstructions and (2) excluding a fraction of the events with the worst predicted uncertainty from the proposed reconstruction framework.
We  have optimized the main parameter $\kappa_{\text{min}}$ of the model that trades off the number of the events considered in the proposed algorithm and the accuracy of the reconstruction.
Finally, the effective value of the estimated CRT was about 494~ps and was only slightly better than the reference time resolution of the PET scanner.

Future work will investigate other aspects of  signal processing by using the proposed statistical model, for instance, the influence of the reference value of the scanner CRT on the effective time resolution.
In this study, the time resolution of the detector system  of 500~ps was assumed.  
However, recent theoretical and experimental studies using small scintillator crystals indicate that the CRT limit is expected at about 100 ps~\cite{Schart2010,Levin2018}. 
In the next steps, we plan to investigate the potential position reconstruction improvement for the systems with CRT smaller than 500 ps.

\bibliographystyle{cs-agh}
\bibliography{references}

\end{document}